\documentclass[10pt,twocolumn,twoside] {IEEEtran}
\usepackage{bbm}
\usepackage{psfrag}
\usepackage{color}
\usepackage[compress]{cite}
\usepackage[pdftex]{graphicx}
\usepackage{amssymb}
\usepackage[cmex10]{amsmath}
\usepackage{amsthm}
\usepackage{algorithm, algorithmic}
\usepackage{array,enumerate}
\usepackage{multirow}
\usepackage[tight, footnotesize]{subfigure}

\newtheorem{definition}{Definition}

\newtheorem{theorem}{Theorem}
\newtheorem{proposition}{Proposition}

\newtheorem{lemma}{Lemma}

\newcommand{\bX}{\mathbf{X}}
\newcommand{\tX}{\tilde{X}}

\renewcommand{\P}{\mathbb{P}}
\newcommand{\Exp}{\mathbf{e}}

\newcommand{\Sus}{\mathbf{s}}
\newcommand{\Inf}{\mathbf{i}}

\newcommand{\Non}{\mathbf{n}}

\newcommand{\cT}{\mathcal{T}}
\newcommand{\cX}{\mathcal{X}}

\newcommand{\IT}{\mathbb{T}}
\newcommand{\Deg}{\textrm{Deg}}

\newcommand{\ofrac}[1]{{\frac{1}{#1}}}

\newcommand{\tc}[1]{^{(#1)}}
\newcommand{\parent}[1]{\mathrm{pa}(#1)}
\newcommand{\children}[1]{\mathrm{ch}(#1)}

\begin{document}
\title{How to Identify an Infection Source with Limited Observations}

\author{
    Wuqiong~Luo,~\IEEEmembership{Student Member,~IEEE}, Wee~Peng~Tay,~\IEEEmembership{Member,~IEEE} and Mei~Leng,~\IEEEmembership{Member,~IEEE}
    \thanks{
        This research was supported by the MOE AcRF Tier 2 Grant MOE2013-T2-2-006. Part of this work was presented at the 1st IEEE Global Conference on Signal and Information Processing, Austin, TX, December 2013.
    }
    \thanks{
        The authors are with the School of Electrical and Electronic Engineering, Nanyang Technological University, 639798, Singapore (e-mail: wluo1@e.ntu.
edu.sg, wptay@ntu.edu.sg, lengmei@ntu.edu.sg).
    }
}
\maketitle

\begin{abstract}
A rumor spreading in a social network or a disease propagating in a community can be modeled as an infection spreading in a network. Finding the infection source is a challenging problem, which is made more difficult in many applications where we have access only to a limited set of observations. We consider the problem of estimating an infection source for a Susceptible-Infected model, in which not all infected nodes can be observed. When the network is a tree, we show that an estimator for the source node associated with the most likely infection path that yields the limited observations is given by a Jordan center, i.e., a node with minimum distance to the set of observed infected nodes. We also propose approximate source estimators for general networks. Simulation results on various synthetic networks and real world networks suggest that our estimators perform better than distance, closeness, and betweenness centrality based heuristics.
\end{abstract}


\section{Introduction}\label{sec:Introduction}
Online social networks like Facebook and Google+ have grown immensely in popularity over the last ten years \cite{Viswanath2009,Kumar2010,Gundotra2013}. For example, Facebook reports over 690 million daily active users on average in its most recent quarterly report, a 27\% increase year-over-year \cite{Facebook2013}. The increase in the size and complexity of such social networks have also been driven by increasing use of technologies like smart phones that facilitate more frequent and faster updates and interactions between members of the social network \cite{Alcarria2010}. Physical social networks consisting of communities of individuals have also grown bigger and more complex due to urbanization and technological advancements in transportation \cite{Kossinets2006}. As a result, a piece of information or rumor posted by one individual in a network can be propagated to a large number of people in a relatively short time. For example, a false rumor about the financial performance of a listed company may be spread by market manipulators to influence the price of the company's stock. An essay containing opinions inciting racial or religious hatred may become widely distributed in an online social network. In some locales like China \cite{Custer2013}, publication of such rumors and opinions are illegal, and law enforcement agencies need to act to identify the rumor and opinion sources. This may be difficult if the source is anonymous, and significant time and effort is expended to trace the IP addresses and identities of the individual profiles carrying or linking to the opinion piece. In another example, it is important to identify the index case of a disease spreading in a community in order to determine the epidemiology of the disease.

In this paper, we are interested to estimate the source of a rumor or disease spreading in a social network. We can think of both the rumor or disease as an ``infection'' that is spreading in the network, and our goal is to find the infection source with only a limited amount of information. For example, given the large size of various online social networks like Facebook, it is impossible for a law enforcement agency to analyze all the profiles that post a particular rumor or link to it. Monitoring of network traffic to identify potential terrorist plots can only be done at select nodes in the network. Similarly, some individuals may have immunity to a disease or may be asymptomatic to it. Finding the infection source is thus a very challenging problem, and we are often limited to knowing just the topology of the network, and a subset of nodes that are infected. However, because of its importance in various applications as described above, the infection source estimation problem has continued to attract considerable interest from the research community over the last two years. In the following, we give a brief overview of works related to infection source estimation.

\subsection{Related Works}\label{subsection:Related_works}

Motivated by applications related to marketing, significant existing works related to infection spreading in a social network have focused on the identification of influential nodes in the network. Each node in a network has a probability of influencing or ``infecting'' its neighbors. Suppose a company wants to promote a new product by initially targeting a few influential nodes in a social network, and hopes these nodes can subsequently influence a large number of nodes in the same network to adopt the new product. The references \cite{Kempe2003,Leskovec2007,Chen2009,Liu2009, Zhang2010,Gomez2010,Bakshy2011} consider the problem of identifying a subset of nodes to maximize the total \emph{expected} influence of the subset, where the expectation is taken over all possible realizations of the infection process. In this paper, we consider a related but different problem. Our aim is to identify a node most likely to be the infection source, given a \emph{particular} realization of the infection process. One of the first works to address the infection source estimation problem is \cite{Shah2011}, which considers a Susceptible-Infected (SI) model, in which there is a single infection source, and susceptible nodes are those with at least one infected neighbor, while infected nodes do not recover. Subsequently, \cite{Luo2013} studies the multiple sources estimation problem under a SI model; \cite{Dong2013} studies the single source estimation problem under a SI model with additional a priori knowledge of the set of suspect nodes; \cite{Zhu2012} considers the single source estimation problem for the Susceptible-Infected-Recovered (SIR) model, where an infected node may recover but can never be infected again; and \cite{Luo2013arxiv} investigates the single source estimation problem for the Susceptible-Infected-Susceptible (SIS) model, where a recovered node is susceptible to be infected again.

All the infection source estimation works listed above assume complete observations of the set of infected nodes, which may not be feasible in a lot of practical scenarios. For example, an user of Google+ may post a rumor on her profile, and choose to make her post public, which can be seen by any Google+ user, or to restrict access of her post to only a select group of friends. The Google+ user who restricts access to her posting will appear to be uninfected to an observer not amongst the select group of friends. The observer will thus only be able to observe a limited subset of all the infected nodes in the network.

The reference \cite{Pinto2012} considers the source estimation problem when only a fraction of infected nodes can be observed. However, \cite{Pinto2012} assumes that for each of these observed nodes, we know the infection time of that node, and from which neighboring node the infection comes from. In an online social network, the posting time of a rumor gives us information about the infection time, but it is often difficult to determine which neighbor the rumor is obtained from, unless the user explicitly references the person she obtained the rumor from. In a physical social network scenario like a disease spreading, the infection times are often not available or inaccurate due to varying degrees of immunity amongst the populace. Therefore, in this paper, we do not make either of the assumptions used by \cite{Pinto2012} so that our proposed methods can be used in more general applications that have limited information. Incorporating knowledge of infection times into our estimation methods may however improve the estimation accuracy for specific applications like online social networks, and is part of our future research work.

We remark that in practice, the probability of exactly detecting the source node can be less than 0.5, depending on the underlying network topology \cite{Shah2011}. For example, suppose that all infected nodes can be observed, and the underlying network is a complete graph. Then, any reasonable source estimator should choose any of the infected nodes with equal probability. Therefore, from a practitioner's point of view, estimating the source node is important only because it allows the practitioner to narrow down the set of potential source nodes to those within a few hops of the estimator. Identifying the correct source in practice requires further domain knowledge and forensic information, a process that can be very time consuming if performed on the full set of infected nodes.

\subsection{Our Contributions}\label{subsection:Contributions}
In this paper, we study the infection source estimation problem for a SI model with limited observations of the set of infected nodes. Our main contributions are the following.
\begin{enumerate}[(i)]
\item For tree networks, we derive an estimator for the source node associated with the most likely infection path that yields the observed subset of infected nodes. We propose an algorithm with time complexity $O(n)$ to find the proposed estimator, where $n$ is the number of nodes in the subtree spanning the set of observed infected nodes.
\item For general networks, we propose an approximate estimator for the source node associated with the most likely infection path. We then convert the problem into a Mixed Integer Quadratically Constrained Quadratic Program (MIQCQP), which can be solved using standard optimization toolboxes. However, since the MIQCQP has high complexity, we also propose a heuristic algorithm with time complexity $O(n^3)$ to find the proposed estimator, where $n$ is the size of the network.
\item  We verify the performance of our estimators on various synthetic tree networks, small-world networks, the western states power grid network of the United States, and part of the Facebook network. In our simulation results, our estimator performs better than the distance, closeness, and betweenness centrality based estimators.
\end{enumerate}

The rest of this paper is organized as follows. In Section \ref{sec:problem_formulation}, we present the system model, assumptions and problem formulation. In Section \ref{sec:single_source_estimation}, we derive a source estimator for tree networks and present an efficient algorithm to find the proposed estimator. In Section \ref{sec:general_network_single_source_estimation}, we derive an approximate source estimator for general networks and suggest two ways to find our proposed estimator. We present simulation results in Section \ref{sec:simulation_results} to verify the effectiveness of the proposed estimators. Finally we conclude and summarize in Section \ref{sec:conclusion}.

\section{Problem Formulation}\label{sec:problem_formulation}

In this section, we first describe our system model and assumptions, and then we provide a problem formulation for finding the infection source based on the most likely infection path. We also summarize various notations and definitions that we use throughout this paper at the end of this section.

\subsection{Infection Spreading Model}
We model a social network using an undirected graph $G=(V,E)$, where $V$ is the set of vertices or nodes that represent the individuals in the network, and $E$ is the set of edges representing relationships between the individuals. Whether an edge exists between two nodes is determined by the application that we are interested in. For example, in identifying a rumor source for an online social network like Facebook or Google+, an edge exists between two individuals if they are friends who can see each other's postings on the network. In determining the index case of a disease spreading in a community, we can create a social network modeling the connections between individuals in the community, where an edge exists between individuals if they are likely to have physical or close proximity contact with each other (e.g., family members living together or colleagues working in the same office). Two nodes are said to be neighbors if there is an edge between them. In this paper, we assume that there is an ``infection'' started by a single node in the graph $G$, and that spreads to neighboring nodes. The term ``infection'' refers to a property possessed by a node, and can mean different things under different contexts. In identifying a rumor source for an online social network, if a node posts the same rumor on its online account, it is considered to be infected. In an actual disease spreading in a community, once a person displays the disease symptoms or has been diagnosed with the disease, he is considered to be infected. We are interested to estimate the source of this infection, given observations of a limited number of nodes in the network.

Models to describe spreading of viral epidemics in human populations have been widely studied in \cite{Bailey1975,Allen1994,Newman2003}, and these have been adapted to model information diffusion in online social networks \cite{Goldenberg2001,Gruhl2004,Domingos2005,Cha2010,Chou2013}. The SI model arises as a natural way to model the spreading process of viral epidemics \cite{Bai2007,Wu2012,Shang2013} and information diffusion \cite{Chou2013}. In the SI model, the nodes have three possible states: \emph{infected} ($\Inf$), \emph{susceptible} ($\Sus$) and \emph{non-susceptible} ($\Non$). Infected nodes are those nodes that possess the infection, and will remain infected throughout. Susceptible nodes are uninfected nodes, but which have at least one infected neighbor. Lastly, non-susceptible nodes are uninfected nodes that do not have any infected neighbors.

In this paper, we adopt a discrete time SI model to describe the infection spreading in $G$. This is an appropriate model for online postings on social networks as most postings are usually not removed, i.e., an infected node stays infected \cite{Gruhl2004}, as well as for modeling opinion dynamics in a social network \cite{Chou2013}, and some disease spreadings \cite{Bai2007,Wu2012}. We note that this by no means models all typical infection spreading processes of interest. Various other researchers, including ourselves, have considered more complex models like continuous time SI models \cite{Shah2011, Luo2013, Dong2013}, SIR models \cite{Zhu2012}, and SIS models \cite{Luo2013arxiv}, but with the assumption that all infected nodes are known. These models are out of the scope of our current paper. We note however that the optimal estimator under our decision framework is the same as that for the more complex models in \cite{Zhu2012, Luo2013arxiv}, leading to the interesting and somewhat surprising result that our proposed estimator is in some sense \textit{universal} (cf. Section \ref{subsec:Most_Likely_Path}).

In our SI model, time is divided into discrete slots, and the state of a node $u$ in time slot $t$ is denoted by a random variable $\bX(u,t)$. At time $t=0$, we assume that there is only one infected node $v^* \in V$, which we call the infection source. We assume that the infection process is a discrete time Markov process with probability measure $\P$, and we assume that a susceptible node becomes infected with probability $p$ at the beginning of the next time slot, where $p \in (0,1)$. We assume that every susceptible node becomes infected independently of each other, while non-susceptible nodes remain uninfected with probability one.

We assume that not all infected nodes exhibit their infected state to an observer. The Google+ example described in Section \ref{subsection:Related_works} is one example. Similarly, individuals who are carriers of a disease may appear to be asymptomatic. An infected node that exhibits its infected status is said to be \emph{explicit}, and we let $\bX(u,t)=\Exp$ when $u$ becomes infected. We let $\bX(u,t)=\Inf$ if $u$ is infected but is non-explicit. A node that is observed to be uninfected can then either be actually uninfected or non-explicit. We call these nodes \emph{non-observable}.

\begin{figure}[!t] 
    \centering
    \psfrag{s}[][][1.1][0]{$\Sus$}
    \psfrag{i}[][][1.1][0]{$\Inf$}
    \psfrag{e}[][][1.1][0]{$\Exp$}
    \psfrag{a}[][][0.9][0]{$1-p$}
    \psfrag{b}[][][0.9][0]{$p(1-q_u)$}
    \psfrag{c}[][][0.9][0]{$pq_u$}
    \psfrag{1}[][][0.9][0]{1}
    \includegraphics[width=0.2\textwidth]{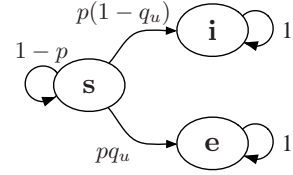}
    \caption{Transition probability of a susceptible node $u$.}
    \label{fig:transition_probability}
\end{figure}

We let $q_u$ be the probability that the node $u$ is explicit, and assume that nodes are explicit or not independently of each other. If $q_u = 1$ for all nodes $u \in V$, our model reduces essentially to that in \cite{Shah2011,Luo2013}, whereas if $q_u$ is close to zero for all nodes $u\in V$, the problem becomes intractable as most nodes appear to be uninfected, making the performance of any estimator of the source node poor in practice. Intuitively, if $q_u \geq p$ for all nodes $u$, then with high probability, we will be able to observe enough number of infected nodes in order to estimate the infection source with reasonably good accuracy relative to that for the SI model where all infected nodes are explicit. However, we only require a weaker assumption in this paper. Throughout this paper, we assume that for all $u \in V$, we have
\begin{align}\label{ineq:q}
\max\left(0,2-\ofrac{p}\right) \leq q_u \leq 1.
\end{align}
(Note that $2-1/p < p$.) In the case where $p > 1/2$, our assumption requires that $q_u$ is sufficiently large for every node $u$. This is because given a sufficiently long amount of time, a large number of nodes will become infected, and if most of these nodes are non-explicit, then it becomes very difficult to find a good estimator for the source. On the other hand, if $p \leq 1/2$, our assumption is trivial and holds automatically. In this case, we note that our model allows us to set $q_u=0$ or $1$ for each node $u$, where those nodes with $q_u=1$ are the ones we actively monitor for the infection. This is similar to \cite{Pinto2012}, except that we do not make the additional assumptions that we know the time an explicit node gets infected, and the neighbor it gets the infection from. The transition probability of a susceptible node is summarized in Fig. \ref{fig:transition_probability}.

Finally, consider the extreme case where the source node has only one neighbor. In this case, the infection can spread in only one direction and any centrality based estimator is expected to perform badly. To avoid this kind of boundary effect, we assume that every node has degree at least two.

\subsection{Most Likely Infection Path and Source Estimator}

Let $\bX^t = \{ \bX(u,\tau): u \in V, 1\leq \tau \leq t\}$ be the collection of the states of all nodes in $V$ from time $1$ to $t$. We say that a realization $X^t= \{X(u,\tau): u \in V, 1\leq \tau \leq t\}$ of $\bX^t$ is an \emph{infection path}. At some elapsed time $t$ since the start of the infection spreading, we observe the set of explicit nodes, which is denoted as $V_\Exp$, and assumed to be non-empty. We say that an infection path $X^t$ is \emph{consistent} with $V_\Exp$ if for all $u \in V_\Exp$, we have $X(u,t) = \Exp$, and no other nodes in $V$ is explicit in $X^t$. We do not assume that we know the elapsed time $t$. Let $\cX_v$ and $\cT_v$ be the set of all possible consistent infection paths and feasible elapsed times respectively, assuming that the infection starts at $v$ and results in $V_{\Exp}$.

\begin{figure}[!t]
  \centering
  \psfrag{A}[][][0.8][0]{$T_{v_3}(v_1;G)$}
  \psfrag{T}[][][0.8][0]{$T_{v_4}(v_1;G)$}
  \psfrag{B}[][][0.9][0]{$V_{\Exp}=\{v_2, v_3\}$}
  \psfrag{G}[][][1.1][0]{$G$}
  \psfrag{1}[][][0.9][0]{$v_1$}
  \psfrag{2}[][][0.9][0]{$v_2$}
  \psfrag{3}[][][0.9][0]{$v_3$}
  \psfrag{4}[][][0.9][0]{$v_4$}
  \psfrag{5}[][][0.9][0]{$v_5$}
  \psfrag{6}[][][0.9][0]{$v_6$}
  \psfrag{7}[][][0.9][0]{$v_7$}
  \psfrag{8}[][][0.9][0]{$v_8$}
  \psfrag{9}[][][0.9][0]{$v_9$}
  \psfrag{c}[][][0.9][0]{$v_{10}$}
  \includegraphics[width=0.35\textwidth]{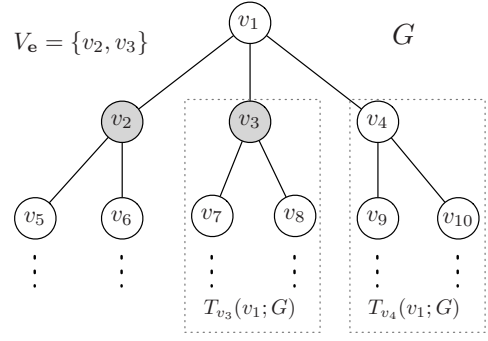}
  \caption[Graph example]{An example network $G$, where shaded nodes are the explicit nodes. Assume that $p=1/2$ and $q_u = 1/2, \ \forall u \in V$. If the elapsed time $t$ is $1$, only $v_1$ can be the source, and the most likely infection path is $X^1(\{v_1, v_2, v_3, v_4\},1)=\{\Inf, \Exp, \Exp, \Sus\}$. The conditional probability of $X^1$ is $(1-q_{v_1})(pq_{v_2})(pq_{v_3})(1-p)=(1/2)^6$. If $t=2$, the possible source nodes are $v_1$, $v_2$, $v_3$ and $v_4$. The probability of the most likely infection path for each of these nodes are $(1/2)^9$, $(1/2)^{10}$, $(1/2)^{10}$ and $(1/2)^{11}$, respectively. It can be shown that if $t > 2$, the infection path probabilities become smaller. Therefore, we see that the most likely elapsed time is $t=1$, the most likely infection path is $X^1$, and $v_1$ is our estimated source.

\ \ Taking $v_1$ as the root, we have $\parent{v_2}=v_1$, and $\children{v_2}=\{v_5, v_6\}$. Since $v_7$ and $v_8$ are one hop away from $v_3$, we have $N_1(v_3;T_{v_3}(v_1;G)) = 2$. The largest distance between $v_1$ and any explicit node is 1, so $\bar{d}(v_1,V_{\Exp})=1$. We also have $T_{v_4}(v_1;G)$ as an example of a non-observable subtree since all nodes in this subtree do not belong to $V_\Exp$.}
\label{fig:numerical_example}
\end{figure}

We want to design a source estimator based on $V_\Exp$ and knowledge of the graph $G$, so as to optimize a statistical criterion. A common approach is to find the maximum likelihood (ML) estimators that maximize the conditional probability of observing the explicit set $V_{\Exp}$, given by
\begin{align*}
(\hat{v}_{ML},\hat{t}_{ML}) \in \arg \max_{\substack{v \in V \\ t \in \cT_v}} \sum_{X^t \in \cX_v} \P(\bX^t = X^t \mid v^* = v).
\end{align*}
It has been shown that finding the ML estimator for a general graph when all nodes are explicit is a \#P-complete problem \cite{Shah2011}. Therefore, finding the ML estimator based on $V_\Exp$ is even more challenging, because unlike the case where all nodes are explicit, we only observe a subset of the infected nodes, which may not contain the infection source. We therefore adopt a different statistical criterion, first proposed by \cite{Zhu2012} to estimate the infection source in a SIR model, and also used by \cite{Luo2013arxiv} in a SIS model. We seek to find the estimator given by
\begin{align}
\hat{v} \in \arg \max_{v \in V} \max_{\substack{t \in \cT_v \\ X^t \in \cX_v}} \P(\bX^t = X^t \mid v^* = v),
\label{equ:proposed_single_source_estimator}
\end{align}
which is the source node associated with a \emph{most likely infection path} out of all possible infection paths that are consistent with $V_\Exp$. See Fig. \ref{fig:numerical_example} for an example of most likely infection path. Note that there may be many possible most likely infection paths, and our aim is to find the source node of any one of them.

In Section \ref{sec:single_source_estimation}, we first find a characterization for a most likely infection path when the underlying graph $G$ is a tree, and then derive a source estimator based on that. For general networks $G$, finding a most likely infection path is difficult. Therefore, in Section \ref{sec:general_network_single_source_estimation}, we propose approximate estimators to identify the source. Extensive simulation results are provided in Section \ref{sec:simulation_results} to verify the performance of our proposed estimators.

\subsection{Some Notations and Definitions}

In this subsection, we list some notations and definitions that we use throughout this paper. See Fig. \ref{fig:numerical_example} for examples of some of these notations and definitions.
\begin{enumerate}[1)]
  \item For any set $A$, we let $|A|$ be the number of elements in $A$. For a graph $H$, we let $|H|$ denote the number of nodes in $H$. Furthermore, for a node $u$, we use $u \in H$ to mean that $u$ is a node in the vertex set of $H$.
  \item Suppose that $G$ is a tree, and $v$ is the root. We assign directions to each edge of $G$ so that all edges point towards $v$. For any $u \in V$, let $\parent{u}$ be the parent node of node $u$ (i.e., the node with an incoming edge from $u$), and $\children{u}$ be the set of child nodes of $u$ in $G$ (i.e., the set of nodes with outgoing edges to $u$).
  \item For any tree $H$ and any node $u$ in $H$, let $N_i(u;H)$ be the set of nodes $i$ hops away from $u$ in $H$.
  \item For any tree $H$ and a pair of nodes $u$ and $v$ in $H$, let $T_u(v;H)$ be the subtree of $H$ rooted at node $u$ with the first link in the path from $u$ to $v$ in $H$ removed.
  \item Given any pair of nodes $v, u \in V$, let $d(v,u)$ denote the length of the shortest path between $v$ and $u$, which is also called the distance between $v$ and $u$. Given any set $A \subset V$, denote the largest distance between $v$ and any node $u \in A$ to be
\begin{align*}
\bar{d}(v,A)=\max_{u \in A} d(v,u).
\end{align*}
We call the largest distance between $v$ and any explicit node $\bar{d}(v,V_{\Exp})$ the \emph{infection range} of $v$. The node with minimum infection range is called a \emph{Jordan center} of $V_{\Exp}$ \cite{Wasserman1994}.
	\item Suppose that $v$ is the infection source. For any node $u$, we say that $T_u(v;G)$ is a \emph{non-observable subtree} if
	\begin{align*}
	T_u(v;G) \bigcap V_{\Exp} = \emptyset.
	\end{align*}
			
	\item For any infection path $X^t$, any subset $J \subset V$, and any $0 \leq i \leq j \leq t$, let $X^t(J,[i,j])$ be the states of nodes in $J$ during time slots $i$ to $j$ in the infection path $X^t$. To avoid cluttered expressions, we abuse notations and let	
	\begin{align}
	&\P\big(\bX^t(J,[i,j])=X^t(J,[i,j]) \mid \nonumber\\
	&\quad \quad \bX^t(J,[i',j'])=X^t(J,[i',j']), v^* =v\big) \nonumber\\
	&= P_v\left(X^t(J,[i,j]) \mid X^t(J,[i',j']) \right). \label{eqn:Pv}
	\end{align}
When we want to remind the reader of the state of a node $u$ at specific times in the conditional probability $P_v(X(u,i) \mid X(u,i'))$, we use the notation $P_v(X(u,i) = a \mid X(u,i') = b)$, where $a, b \in \{\Inf,\Sus,\Non,\Exp\}$ are the states of $u$ at times $i$ and $i'$ respectively.
	
	\item For any $v\in V$ and any feasible elapsed time $t \in \cT_v$, we say that an infection path $X^t$ is most likely for $(v,t)$ if $X^t \in \arg\max_{\tX^t \in \cX_v} P_v(\tX^t)$. We say that an infection path $X^t$ is most likely for $v$ if $(X^t, t) \in \arg\max_{t'\in \cT_v, \tX^{t'} \in \cX_v} P_v(\tX^{t'})$. Finally, an infection path $X^t$ is called a most likely infection path if there exists some $v\in V$ and $t \in \cT_v$ such that $P_v(X^t) = \max_{u\in V, t'\in \cT_u, \tX^{t'} \in \cX_u} P_u(\tX^{t'})$.
\end{enumerate}

\section{Source estimation for trees} \label{sec:single_source_estimation}

In this section, we consider the case where the underlying network $G$ is a tree. We first derive some properties of a most likely infection path, and then show that the source estimator associated with the most likely infection path that we have characterized is given by the Jordan center of $V_\Exp$.

\subsection{A Most Likely Infection Path}

In this subsection, we show that although we have assumed that $G$ is an infinite tree, we can restrict our search for a most likely infection path for $v$ to the subgraph of $G$ with nodes within the infection range $\bar{d}(v,V_{\Exp})$ of $v$. In the following lemma, we first show that for any source node $v\in V$, a most likely infection path for $v$ with a finite number of infected nodes can be found. It's proof is provided in Appendix \ref{appendix:lemma:empty_subtree}.

\begin{lemma}\label{lemma:empty_subtree}
Suppose that $G$ is a tree. Then, for any node $v\in V$, and feasible elapsed time $t\in \cT_v$, there exists a most likely infection path $X^t$ for $(v,t)$ such that for any $u \in V$ with non-observable $T_u(v;G)$, we have $X(u,\tau) \ne \Inf$ for all $\tau \leq t$.
\end{lemma}

The following lemma, whose proof is given in Appendix \ref{appendix:lemma:better_later}, shows that given the elapsed time $t$, a most likely infection path for a node $v$ is given by a path whose nodes ``resist'' the infection, and each node becomes infected only at the latest possible time.

\begin{lemma}\label{lemma:better_later}
Suppose that $G$ is a tree, and the infection source is $v\in V$. Let $H$ be the minimum connected subgraph of $G$ that spans $V_{\Exp}$ and $v$. Suppose that the elapsed time is $t \in \cT_v$. Then, for any $u \in H \backslash \{v\}$, the first infection time $t_u$ for $u$ in any infection path is bounded by
\begin{align}
d(v,u)\leq t_u \leq t-\bar{d}(u, T_u(v;H)). \label{equ:first_infection_time_bound}
\end{align}
Furthermore, there exists a most likely infection path $X^t$ for $(v,t)$ such that the first infection time for $u \in H \backslash \{v\}$ is given by
\begin{align}
t_u = t-\bar{d}(u, T_u(v;H)). \label{equ:optimal_first_infection_time}
\end{align}
\end{lemma}

Lemma \ref{lemma:empty_subtree} and Lemma \ref{lemma:better_later} characterize a most likely infection path consistent with $V_\Exp$, with the property that a minimum number of nodes are infected, and each infected node becomes infected at the latest possible time. We call this \emph{the latest infection path}.
\begin{definition}\label{def:latest_path}
Suppose that $G$ is a tree. For any $v\in V$, and any feasible elapsed time $t \in \cT_v$, the latest infection path $X^t$ for $(v,t)$ is the infection path that satisfies the following properties:
\begin{enumerate}[(i)]
	\item Let $H$ be the minimum connected subtree of $G$ that spans $V_\Exp \cup \{v\}$. Then, $X^t(u,\tau) \in \{\Sus,\Non\}$ for all $u \notin H$ and for all $\tau \leq t$.
	\item For each $u \in H \backslash\{v\}$, the first infection time of $u$ is $t_u = t-\bar{d}(u, T_u(v;H))$.
\end{enumerate}
\end{definition}
The following proposition then follows from Lemma \ref{lemma:empty_subtree} and Lemma \ref{lemma:better_later}.
\begin{proposition}\label{prop:latest_path}
Suppose that $G$ is a tree. Then for any $v\in V$, and any feasible elapsed time $t \in \cT_v$, the latest infection path for $(v,t)$ is a most likely infection path for $(v,t)$.
\end{proposition}

We now show that a most likely infection path for any $v\in V$ is the latest infection path for $(v,t)$, where $t$ is chosen to be as small as possible.

\begin{proposition}\label{prop:optimal_t} Suppose that $G$ is a tree. For any $v\in V$, we have the following.
\begin{enumerate}[(a)]
  \item \label{prop:optimal_t_feasible_set} The set of all feasible elapsed times is $\cT_v=[\bar{d}(v,V_{\Exp}), \infty)$.
  \item \label{prop:optimal_t_monotonically_decreasing} For the sequence of latest infection paths $\{X^{t}\}_{t\in \cT_v}$, we have $P_v(X^t)$ is monotonically decreasing in $t \in \cT_v$.
  \item \label{prop:optimal_t_optimal_t} The most likely elapsed time is given by $\bar{d}(v,V_{\Exp})$, and a most likely infection path for $v$ is the latest infection path for $(v,\bar{d}(v,V_{\Exp}))$.
\end{enumerate}
\end{proposition}

\begin{IEEEproof}
Claim \eqref{prop:optimal_t_feasible_set} follows because the infection can propagate at most one hop further from the source node $v$ in one time slot, therefore if $t<\bar{d}(v,V_{\Exp})$, the infection can not reach the explicit nodes $\bar{d}(v,V_{\Exp})$ hops away from $v$.

Next, we show claim \eqref{prop:optimal_t_monotonically_decreasing}. Fix a $t \in \cT_v$ and consider the latest infection paths $X^{t}$ and $X^{t+1}$. Let $H$ be the minimum connected subtree of $G$ that spans $V_{\Exp}$ and $v$. Suppose that $H \ne \{v\}$, then from Definition \ref{def:latest_path}, we have $X^{t+1}(V,[2,t+1]) = X^{t}(V,[1,t])$ and $X^{t+1}(w,1) = \Sus$ for all $w \in N_1(v;H)$, yielding
\begin{align*}
\frac{P_v(X^{t+1})}{P_v(X^{t})} \leq & \prod_{w \in N_1(v;H)}P_v(X^{t+1}(w,1) = \Sus \mid X^{t+1}(w,0) = \Sus)
\\
= & (1-p)^{|N_1(v;H)|} < 1,
\end{align*}
where the last inequality follows because $H \ne \{v\}$ and $|N_1(v;H)| > 1$. On the other hand if $H = \{v\}$, the infection does not spread from $v$ in both latest paths $X^t$ and $X^{t+1}$, and we have
\begin{align*}
\frac{P_v(X^{t+1})}{P_v(X^{t})} &= (1-p)^{|N_1(v;G)|} < 1,
\end{align*}
since all neighbors of $v$ remain susceptible throughout the elapsed time, and we have assumed that the degree of $v$ is at least two. This proves claim \eqref{prop:optimal_t_monotonically_decreasing}. Claim \eqref{prop:optimal_t_optimal_t} now follows from claim \eqref{prop:optimal_t_feasible_set} and claim \eqref{prop:optimal_t_monotonically_decreasing}, and the proof of the lemma is complete.
\end{IEEEproof}

\subsection{Source Associated with a Most Likely Infection Path}\label{subsec:Most_Likely_Path}

In this subsection, we derive the source estimator associated with a most likely infection path. We first show that we can find an infection path for a node with a smaller infection range that is more likely than any most likely infection path of another node with a larger infection range. This in turn implies that the source estimator we are looking for is the Jordan center of $V_\Exp$. We start with two lemmas that show the relationship between the latest infection paths of two neighboring nodes.
\begin{lemma}\label{lemma:tv=tu-1}
Suppose that $G$ is a tree, and let $H$ be the minimum connected subtree of $G$ spanning $V_{\Exp}$. Suppose that $u$ and $v$ are neighboring nodes in $H$ with $\bar{d}(v,V_\Exp) < \bar{d}(u,V_\Exp)$. Then, we have
\begin{enumerate}[(i)]
  \item \label{prop:tv=tu-1_leaf} $l \in T_v(u;H)$, for all $l \in \arg \max_{x\in V_{\Exp}}d(u,x)$; and
  \item \label{prop:tv=tu-1_tv=tu-1} $\bar{d}(v,V_\Exp)=\bar{d}(u,V_\Exp)-1$, and there exists $l\in T_v(u;H)$ such that $d(v,l) = \bar{d}(v,V_\Exp)$.
\end{enumerate}
\end{lemma}
\begin{IEEEproof}
To prove \eqref{prop:tv=tu-1_leaf}, we note that if $l \notin T_v(u; H)$, we have $\bar{d}(v,V_{\Exp})\geq d(v,l)=d(u,l)+1=\bar{d}(u,V_\Exp)+1$, a contradiction. Therefore, \eqref{prop:tv=tu-1_leaf} holds. Then for $l$ such that $d(u,l)=\bar{d}(u,V_\Exp)$, we have $\bar{d}(v,V_{\Exp})\geq d(v,l)=d(u,l)-1=\bar{d}(u,V_\Exp)-1$ since $l \in T_v(u;H)$. This implies \eqref{prop:tv=tu-1_tv=tu-1}, and the lemma is proved.
\end{IEEEproof}

\begin{lemma}\label{lemma:better_neighbor}
Suppose that $G$ is a tree, and let $H$ be the minimum connected subtree of $G$ spanning $V_{\Exp}$. Then, for any pair of neighboring nodes $u$ and $v$ in H with  $d_v=\bar{d}(v,V_\Exp) < d_u=\bar{d}(u,V_\Exp)$, we have
\begin{align*}
P_v(X^{d_v}) > P_u(X^{d_u}),
\end{align*}
where $X^{d_v}$ and $X^{d_u}$ are the latest infection paths for $(v,d_v)$ and $(u,d_u)$ respectively.
\end{lemma}
\begin{IEEEproof}
To prove the lemma, it suffices to construct an infection path $\tX^{d_v}$ with source node $v$, and show that it has higher conditional probability than $X^{d_u}$. Let $t_v$ be the first infection time of node $v$ in the infection path $X^{d_u}$. We first show that $t_v=1$. Since $u$ is the infection source, the infection can propagate at most $d_u-t_v$ hops away from node $v$ within the subtree $T_v(u;H)$. From Lemma \ref{lemma:tv=tu-1}(\ref{prop:tv=tu-1_tv=tu-1}), if $t_v>1$, we have $d_v=d_u-1>d_u-t_v$, a contradiction. Therefore, we must have $t_v=1$ in the infection path $X^{d_u}$. Let $\tX^{d_v}(T_v(u;H), [1, d_v]) = X^{d_u}(T_v(u;H), [2, d_u])$, and we have
\begin{align}
\frac{P_u(X^{d_u}(T_v(u;H), [1,d_u]))}{P_v(\tilde{X}^{d_v}(T_v(u;H),[1,d_v]))}
&= p, \label{equ:better_neighbor_subtree_v}
\end{align}
where the equality holds because the probability that $v$ is explicit or not appears in both the numerator and denominator.

Consider any node $w \in N_1(u, T_u(v;H))$. Since $d(v,w) = 2$, it takes at least two time slots for an infection starting at $v$ to reach $w$. Moreover, since $d_u = d_v+1$, by Lemma \ref{lemma:better_later} and Lemma \ref{lemma:tv=tu-1}\eqref{prop:tv=tu-1_leaf}, the first infection time of $w$ in the path $X^{d_u}$ is at least 3. Therefore, we can set $\tX^{d_v}(T_u(v;H), [1, d_v])=X^{d_u}(T_u(v;H), [2, d_u])$, and we obtain
\begin{align}
\frac{P_u(X^{d_u}(T_u(v;H), [1,d_u]))}{P_v(\tilde{X}^{d_v}(T_u(v;H),[1,d_v]))}
&= \frac{1}{p}{(1-p)^{2\left| N_1(u, T_u(v;H))\right|}} \nonumber \\
&< \frac{1}{p}, \label{equ:better_neighbor_subtree_u}
\end{align}
where the inequality follows by the assumption that every node has degree at least two. Multiplying \eqref{equ:better_neighbor_subtree_v} by \eqref{equ:better_neighbor_subtree_u}, we obtain
\begin{align}
&\frac{P_u(X^{d_u}(H,[1,d_u]))}{P_v(\tilde{X}^{d_v}(H, [1,d_v]))}
<p \cdot \frac{1}{p} = 1. \label{equ:better_neighbor_H}
\end{align}

Finally, we consider the nodes in $G\backslash H$. Let $T_z(v;G)$ be any non-observable subtree such that $z \in G \backslash H$ has a neighboring node $w \in H$. By Lemma \ref{lemma:empty_subtree}, $z$ remains uninfected in $X^{d_u}$. Let $z$ stay uninfected in $\tilde{X}^{d_v}$ as well. Let the node $w$ first become infected at time $t_w(u)$ in $X^{d_u}$ and at time $t_w(v)$ in $\tX^{d_v}$.
Then, $z$ stays uninfected in $X^{d_u}$ and $\tX^{d_v}$ with probabilities $(1-p)^{d_u - t_w(u)}$ and $(1-p)^{d_v - t_w(v)}$, respectively.
Since $d_v = d_u-1$ and $t_w(v) \geq t_w(u) - 1$,
we have $d_u-t_w(u) \geq d_v-t_w(v)$, and $P_u(X^{d_u}(G\backslash H, [1,d_u])) \leq P_v(\tilde{X}^{d_v}(G\backslash H, [1,d_v]))$.
Combining this with \eqref{equ:better_neighbor_H}, we conclude that $P_u(X^{d_u}) < P_v(\tX^{d_v})$, and the proof is complete.
\end{IEEEproof}

We are finally ready to show that the Jordan centers of $V_\Exp$ are the source estimators in \eqref{equ:proposed_single_source_estimator}.

\begin{theorem}\label{theorem:single_source_estimate_Jordan_infection_center}
Suppose that $G$ is a tree, then the source estimator in \eqref{equ:proposed_single_source_estimator} associated with the most likely infection path is a Jordan center of $V_\Exp$, given by
\begin{align}
\hat{s} \in \arg \min_{v \in V} \bar{d}(v,V_{\Exp}). \label{equ:Jordan_infection_center}
\end{align}
\end{theorem}

\begin{IEEEproof}
It can be shown that if $G$ is a tree, then there are at most two Jordan centers for $V_{\Exp}$, and if there are indeed two Jordan infection centers, they are neighboring nodes \cite{Zhu2012}. If there are two neighboring Jordan centers, we can treat them as a single virtual node, therefore without loss of generality, we assume that there is only one Jordan center $\hat{s}$. Let $H$ be the minimum connected subtree of $G$ spanning $V_{\Exp}$. Then $\hat{s} \in H$. Consider any path $(\hat{s}, v_1, v_2, \cdots, v_m)$ in $H$, where $m \geq 1$. We show that
\begin{align}
\bar{d}(\hat{s},V_\Exp) < \bar{d}(v_1,V_\Exp) < \ldots < \bar{d}(v_m,V_\Exp).\label{increasingd}
\end{align}
The first inequality in \eqref{increasingd} holds by assumption. We now show that the rest of the inequalities in \eqref{increasingd} also hold. Choose a $l \in V_\Exp$ such that $d(\hat{s},l)=\bar{d}(\hat{s},V_{\Exp})$. If $l \notin T_{v_1}(\hat{s};H)$, we have $\bar{d}(v_i,V_{\Exp})=d(\hat{s},l)+i$ for $1 \leq i \leq m$, so \eqref{increasingd} holds. Suppose now that $l \in T_{v_1}(\hat{s};H)$. Note that the set $A=V_{\Exp} \backslash T_{v_1}(\hat{s};H)$ is non-empty, otherwise $\bar{d}(v_1,V_\Exp) < \bar{d}(\hat{s},V_\Exp)$ and $\hat{s}$ cannot be a Jordan center. Consider a node $l'$ such that $l'=\arg \max_{v\in A} d(\hat{s},v)$. If $d(\hat{s},l') \leq d(\hat{s},l)-2$, we have
\begin{align*}
\bar{d}(v_1,V_{\Exp}) &= \max \left(d(v_1,l'),d(v_1,l)\right) \\
&=\max \left(d(\hat{s},l')+1,d(\hat{s},l)-1\right) \\
&=d(\hat{s},l)-1,
\end{align*}
and the infection range of $v_1$ is less than that of $\hat{s}$, a contradiction. Therefore, we have $d(\hat{s},l') \geq d(\hat{s},l)-1$. Suppose that $\bar{d}(v_{i+1},V_\Exp) \leq \bar{d}(v_{i},V_\Exp)$ for some $i \in [1,m-1]$. Let $\tilde{l}$ be a node such that $d(v_i, \tilde{l}) = \bar{d}(v_{i},V_\Exp)$. Then, we must have $\tilde{l} \in T_{v_{i+1}}(v_i;H)$, otherwise we have a contradiction. We then have
\begin{align*}
\bar{d}(v_{i+1},V_\Exp)
&\geq d(v_{i+1},l') \\
&= d(\hat{s},l') + i + 1\\
&\geq i+d(\hat{s},l)\\
&\geq 2i + d(v_i,\tilde{l}) \\
& > d(v_i,\tilde{l}),
\end{align*}
a contradiction. Therefore \eqref{increasingd} holds. By repeatedly applying Lemma \ref{lemma:better_neighbor}, Proposition \ref{prop:latest_path} and Proposition \ref{prop:optimal_t}(\ref{prop:optimal_t_optimal_t}), we have that any most likely infection path for $\hat{s}$ has higher probability than that for $v_m$, and the theorem is proved.
\end{IEEEproof}

We observe that our source estimator is the same as that for the SIR infection process \cite{Zhu2012} and the SIS infection process \cite{Luo2013arxiv}. This is somewhat surprising as the spreading models are significantly different from each other. (Note also that our proofs differ significantly from that in \cite{Zhu2012}.) This indicates that the statistical criterion in \eqref{equ:proposed_single_source_estimator} is robust to the underlying infection model, and the Jordan center is a universal source estimator.

\subsection{Finding a Jordan Center}
A centralized linear time complexity algorithm has been proposed in \cite{Hedetniemi1981} to find the Jordan center in a tree. It first computes the diameter of the tree and then returns a midpoint of any longest path in the tree as the Jordan center. In this subsection, we present a message passing algorithm, somewhat similar to that of \cite{Hedetniemi1981} in the quantities being computed at each node, but which can be implemented in a distributed fashion. Our algorithm also has linear time complexity.

Let $H$ be the minimum connected subtree of $G$ spanning $V_{\Exp}$. We assume that $|H|>2$ since otherwise finding the Jordan center is trivial. Our proposed Jordan Center Estimation (JCE) algorithm is formally presented in Algorithm \ref{algo:JCE}. The main idea behind the algorithm is that for $|H|>2$, a Jordan center $v$ must satisfy the following necessary and sufficient conditions: (i) it has degree at least 2 in $H$, and (ii) if $M = \{\rho_u : u \in N_1(v;H),\  \rho_u$ is a path with the maximum length among all paths with first edge being $(v,u)\}$, then the difference in lengths of the longest and second longest paths in $M$ is at most 1. This can be shown using the same arguments as that in the proof of Theorem \ref{theorem:single_source_estimate_Jordan_infection_center}.

JCE first randomly chooses a non-leaf node $r \in H$ as the root node. It then performs an Upward Message-passing procedure, starting from the leaf nodes up to the root $r$, where the message passed from a node $v$ to its parent node $\parent{v}$ consists of its own identity and the length of the longest path in $T_v(r;H)$.

This upward message-passing procedure terminates when the root receives all messages from its child nodes. The details of the upward message-passing procedure are shown in lines \ref{JCE:Upward_start} to \ref{JCE:Upward_end} in Algorithm \ref{algo:JCE}. Since each node only passes one message to its parent, the overall complexity of the upward message-passing procedure is $O(|H|)$.

In the Downward Message-passing procedure, the root node $r$ first identifies the two paths in $M$ with the longest lengths $\ell_{1}(r)$ and $\ell_{2}(r)$. If $\ell_{1}(v)-\ell_{2}(v) \leq 1$, JCE returns $r$ as the Jordan center. Otherwise, it computes a message $g_{r}(r\tc{1}) = \ell_{2}(r)+1$ and sends to $r\tc{1}$, the child node with the longest path in the Upward Message-passing procedure. The same process is repeated until a leaf node is reached. The details are presented in lines \ref{JCE:Downward_start} to \ref{JCE:Downward_end} in Algorithm \ref{algo:JCE}. The complexity of the downward message-passing process is bounded by the diameter of $H$. As a result, the overall complexity of JCE is $O(|H|)$.

\begin{algorithm}[!t]
\caption{Jordan Center Estimation (JCE) Algorithm}
\label{algo:JCE}
\begin{algorithmic}[1]
\STATE{\textbf{Input}: $H$ is the minimum connected subtree of $G$ spanning $V_\Exp$, with $|H|>2$.}
\STATE{\textbf{Output}: $\hat{s}$, the Jordan center for $V_\Exp$.}
\STATE{\textbf{Initialization}: randomly select a non-leaf node $r \in H$ as the root node}
\STATE{\textbf{Upward Message-passing:}}\label{JCE:Upward_start}
\FOR{each $v \in H$}
    \IF{$v$ is a leaf}
        \STATE{$f_v(\parent{v})=1$}
    \ELSE
        \STATE{Store $v\tc{1} =\arg\max_{u \in \children{v}} f_{u}(v)$, $\ell_1(v)=f_{v\tc{1}}(v)$, and $\ell_{2}(v)=\max_{u \in C} f_{u}(v)$, where $C=\children{v}\backslash \{v\tc{1}\}$, with $\ell_{2}(v)=0$ if $C = \emptyset$. Ties are broken randomly.}
        \STATE{$f_v(\parent{v})=\ell_{1}(v)+1$}
    \ENDIF
    \STATE{Pass $f_v(\parent{v})$ and its identity to $\parent{v}$}
\ENDFOR\label{JCE:Upward_end}
\STATE{\textbf{Downward Message-passing:}}\label{JCE:Downward_start}
\FOR{each $v \in H$ starting from root $r$}
    \IF{$v$ is not the root $r$}
        \STATE{$\ell_{2}(v)=\max(\ell_{2}(v), g_{\parent{v}}(v))$}
    \ENDIF
    \IF{$\ell_{1}(v)-\ell_{2}(v) \leq 1$}
        \STATE{$\hat{s}=v$}
    \ELSE
        \STATE{Pass $g_v(v\tc{1})=\ell_{2}(v)+1$ to $v\tc{1}$}
    \ENDIF
\ENDFOR\label{JCE:Downward_end}
\RETURN{$\hat{s}$}
\end{algorithmic}
\end{algorithm}

\section{Source Estimation for General Networks} \label{sec:general_network_single_source_estimation}

In this section, we derive an approximate source estimator for the case where the underlying network $G$ is a general network. We also suggest heuristic algorithms to find our proposed source estimator.

Suppose that the neighbor from which a susceptible node obtains its infection is randomly chosen from one of its infected neighbors. Then, the path traced out by an infection spreading in $G$ is a tree. For any infection path $X^t$, let $T(X^t)$ be the subtree of $G$ traced out by $X^t$. Any tree $T_v$ with root $v$, for which there exists an infection path consistent with $V_\Exp$ is said to be an infection tree consistent with $V_\Exp$. Let $\IT_v$ be the set of infection trees consistent with $V_\Exp$, and have source node $v$. Then, we have
\begin{align}
\max_{v \in V} \max_{\substack{t \in \cT_v \\ X^t \in \cX_v}} P_v(X^t)
&= \max_{v \in V}\max_{T \in \IT_v} \max_{\substack{t \in \cT_v \\ X^t: T(X^t) = T}} P_v(X^t), \label{max_tree}
\end{align}
which implies that to find a source node associated with the most likely infection path in \eqref{equ:proposed_single_source_estimator}, we first find the most likely infection tree consistent with $V_\Exp$, and then find a most likely infection path that traces out this infection tree using the results in Section \ref{subsec:Most_Likely_Path}. However, finding the set of infection trees consistent with $V_\Exp$ is difficult. We derive a simple property that $\IT_v$ must satisfy for each $v$, suggest an approximation for it, and provide two heuristic methods to find the approximate most likely infection tree.

\begin{lemma}\label{lemma:most_likely_infection_tree_property}
Suppose that $G$ is a general network, and $v \in V$ is the infection source. Then, there is no loss in optimality in \eqref{max_tree} if we restrict $\IT_v$ to be the set of all infection trees consistent with $V_\Exp$ that have all non-source leaf nodes in $V_\Exp$.
\end{lemma}
\begin{IEEEproof}
The proof follows from Lemma \ref{lemma:empty_subtree} because any non-observable leaf node is the root of a non-observable subtree.
\end{IEEEproof}
From Lemma \ref{lemma:most_likely_infection_tree_property} and the fact that the source estimator for a tree is given by a Jordan center, we construct a subgraph $H_v$ of $G$ for each $v\in V$, by first finding a shortest path tree from $v$ to each node in $V_\Exp$, and then adding all edges in $G$ incident to the nodes of the shortest path tree. We approximate $\IT_v$ by the set of spanning trees of $H_v$, denoted as $\hat{\IT}_v$, and adopt the approximate source estimator given by
\begin{align}
\tilde{s}
&= \max_{v \in V}\max_{T \in \hat{\IT}_v} \max_{\substack{t \in \cT_v \\ X^t: T(X^t) = T}} P_v(X^t). \label{approx_source_general}
\end{align}
We now turn to finding the most likely infection tree in the set $\hat\IT_v$. We start with a characterization of its probability in the following result. For any source node $v$, and any infection tree $T \in \hat\IT_v$, let $D_T(u) = \bar{d}(u,T_u(v;T))$ be the height of the subtree of $T$ rooted at $u$.

\begin{lemma}\label{lemma:most_likely_infection_path_for_infection_tree}
Suppose that $G$ is a general network, and $v \in V$ is the infection source. Then, for any infection tree $T$ with root $v$ and consistent with $V_\Exp$, we have
\begin{align}
&\max_{\substack{t \in \cT_v \\ X^t: T(X^t) = T}} P_v(X^t) \nonumber  \\
&= p^{|T|-1} (1-p)^{\sum_{u \in T \backslash \{v\}} (D_T(\parent{u})-D_T(u))-|T|+1} \nonumber\\
&\quad \cdot \prod_{u \in V_{\Exp}}q_u \prod_{u \in T \backslash V_{\Exp}} (1-q_u). \label{equ:most_likely_infection_path_pa}
\end{align}
\end{lemma}
\begin{IEEEproof}
Every non-source node $u \in T \backslash \{v\}$ gets infected during the infection spreading process, and this occurs with probability $p^{|T|-1}$. From Lemma \ref{lemma:better_later} and Proposition \ref{prop:optimal_t}, in a most likely infection path for $v$, the first infection time for any node $u \in T$ is $\tilde{t}_u = d-D_T(u)$, where $d = \bar{d}(v,V_\Exp)$. The parent node $\parent{u}$ gets infected at time $\tilde{t}_{\parent{u}} = d-D_{\parent{u}}$, therefore there are $\tilde{t}_u-\tilde{t}_{\parent{u}}-1 = D_T(\parent{u}) - D_T(u)-1$ time slots in which that $u$ remains susceptible, and this occurs with probability $(1-p)^{D_T(\parent{u})-D_T(u)-1}$. Taking the product over all non-source nodes in $\IT_v$, we obtain the lemma.
\end{IEEEproof}

From Lemma \ref{lemma:most_likely_infection_path_for_infection_tree}, we see that our proposed estimator in \eqref{approx_source_general} can be found by finding the node $v$ that maximizes
\begin{align*}
|T^*_v|\log p + (F^*_v - |T^*_v|)\log(1-p) + \sum_{u \in T^*_v \backslash V_{\Exp}}\log(1-q_u),
\end{align*}
where
\begin{align}\label{equ:objective_function_pa}
F^*_v &= \min_{T \in \hat\IT_v}\sum_{u \in T\backslash \{v\}}\left(D_T(\parent{u})-D_T(u)\right),
\end{align}
and $T^*_v$ is the minimizer in the right hand side of \eqref{equ:objective_function_pa}.

In the following, we propose two heuristic methods to find $T^*_v$ and $F^*_v$ for any $v \in V$. Although we have assumed that $G$ is infinite in Section \ref{sec:problem_formulation} to simplify our theoretical analysis, in practice, we have access to only a finite graph $G$. We then iterate either of our proposed methods over all nodes in $G$.

\subsection{MIQCQP}\label{subsection: MIQCQP}

For any $v \in V$, let $H_v$ be the subgraph of $G$ used in defining $\hat\IT_v$ (cf.\ discussion preceding \eqref{approx_source_general}). By expanding the telescoping sum in \eqref{equ:objective_function_pa}, it can be shown that the objective function in \eqref{equ:objective_function_pa} is equivalent to the objective function of the following MIQCQP,
\begin{align}
\min &\sum_{i,j \in H_v}E_{ij}D_i - \sum_{i \in H_v \backslash \{v\}} D_i \label{equ:MIQCQP_objective} \\
\text{subject to } & \sum_{j \in H_v}E_{ji} =
\begin{cases}
    0, & \text{if $i = v$}\\
    1, & \text{if $i \ne v$}
\end{cases} \label{equ:MIQCQP_constraint_incoming_edges}\\
& \sum_{i,j \in H_v} E_{ij} = |H_v|-1 \label{equ:MIQCQP_constraint_totoal_edges}\\
& E_{ij} \in \{0,1\}, \ \ \forall i,j \in H_v \label{equ:MIQCQP_constraint_integer}\\
& D_i \geq (D_j + 1)E_{ij}, \ \ \forall i \in H_v, \forall j \in N_1(i;H_v) \label{equ:MIQCQP_constraint_D_lower_bound}\\
& D_i \leq \sum_{j \in N_1(i;H_v)} (D_j+1)E_{ij},  \ \ \forall i \in H_v\label{equ:MIQCQP_constraint_D_upper_bound}
\end{align}
where $E_{ij} = 1$ if and only if the edge $(i,j)$ is chosen as part of $T^*_v$, and $D_i$ is a variable corresponding to $D_{T^*_v}(i)$.

The above MIQCQP does not find the exact optimal $T^*_v$ in \eqref{equ:objective_function_pa} because of several approximations that we have made. The constraints \eqref{equ:MIQCQP_constraint_incoming_edges}-\eqref{equ:MIQCQP_constraint_integer} restrict the feasible solution space to the spanning trees of $H_v$. However, it is not possible represent the quantities $D_T(u)$ in \eqref{equ:objective_function_pa} exactly as quadratic constraints. We use the quadratic constraints \eqref{equ:MIQCQP_constraint_D_lower_bound}-\eqref{equ:MIQCQP_constraint_D_upper_bound} as approximations. The constraint \eqref{equ:MIQCQP_constraint_D_lower_bound} derives from the fact that for all nodes $i$ in a tree $T$ spanning $H_v$, we have $D_T(i) \geq D_T(j)+1$ for all $j$ that are child nodes of $i$ in $T$. In addition, for a non-source node $i$ that has only one child node $j$, the value $D_i$ is canceled out in \eqref{equ:MIQCQP_objective}. In order to improve the convergence speed, we introduce the quadratic constraint \eqref{equ:MIQCQP_constraint_D_upper_bound} as an upper bound for $D_i$.

The MIQCQP in \eqref{equ:MIQCQP_objective} can be solved using the OPTI Toolbox \cite{Currie2012}, which utilizes Solving Constraint Integer Programs (SCIP) \cite{Achterberg2009} as the solvers. However, these solvers have high complexity, and are not suitable for large networks. Therefore, we propose an alternative, low-complexity heuristic algorithm in the following and compare their performances in Section \ref{sec:simulation_results}.

\subsection{Reverse Greedy}\label{subsection:reverse_greedy}
It can be shown that
\begin{align}
F^*_v = &\min_{T\in \IT_v}\sum_{u \in T} (\Deg_T(u)-2)D_T(u) + 2D_T(v),  \label{equ:objective_function_deg}
\end{align}
where $\Deg_T(u)$ is the number of neighboring nodes of $u$ in $T$.
We see from \eqref{equ:objective_function_deg} that to obtain the minimizer, the degree of $u$ should be chosen to be small if $D_T(u)$ is large. Based on this intuition, we propose a heuristic algorithm in Algorithm \ref{algo:RG}, which we call the Reverse Greedy (RG) algorithm. The algorithm attempts to adjust the infection tree found so that those nodes close to the source node has lower degrees, while those further away have correspondingly higher degrees.

Given a node $v \in V$, RG first constructs a shortest path tree $T$ rooted at $v$ and spanning $H_v$ using a breadth first search algorithm \cite{Cormen2001}, incurring a time complexity of $O(|H_v|)$. Starting with a leaf node $x$ that is furthest away from $v$, RG adjusts the tree $T$ by choosing a neighboring node $y$ of $x$ with the largest $d(v,y)$ value (with ties broken by another criterion given in lines \ref{algo:RG:select_y_start} to \ref{algo:RG:select_y_end} in Algorithm \ref{algo:RG}), and attaching $x$ to $y$. This procedure is repeated up the tree $T$ until the node $v$ is reached. At each node $x$, the number of neighbors is $O(|H_v|)$ and lines \ref{algo:RG:update_start} to \ref{algo:RG:update_end} has a time complexity of $O(|H_v|)$. Therefore, the complexity of RG is $O(|H_v|^2)$. Since we need to iterate over all nodes in order to find the source estimator, the overall complexity is $O(|V|^3)$.

\begin{algorithm}[!ht]
\caption{Reverse Greedy (RG) Algorithm}
\label{algo:RG}
\begin{algorithmic}[1]
\STATE{\textbf{Inputs}: $v\in V$, and $H_v$ (cf.\ discussion preceding \eqref{approx_source_general})}
\STATE{\textbf{Outputs}: $F^*_v$ in \eqref{equ:objective_function_deg} and its minimizer $T^*_v$}.
\STATE{Construct a shortest path tree $T$ rooted at $v$ and that spans $H_v$ using breadth first search \cite{Cormen2001}. Compute $D_T(u)$ and $U_T(u) = d(v,u)$ for all $u \in T$.}
\FOR{$d = D_T(v)$ DownTo 1}
    \FOR{each node $x \in \{u: U_T(u) = d\}$, in order of increasing $D_T(x)$}
    \STATE{Set $Y=\emptyset$}\label{algo:RG:select_y_start}
        \FOR{each neighboring node $y$ of $x$ in $H_v$}
            \IF{$U_T(y) \le D_T(v)-D_T(x)-1$}
                \STATE{Add $y$ into set $Y$}
                \IF{$y$ is the parent of $x$ in $T$}
                    \STATE{Set $D'_T(y)$ to be $\bar{d}(y, T_y(v;T)\backslash T_x(v;T))$}
                \ELSE
                    \STATE{Set $D'_T(y) = D_T(y)$}
                \ENDIF
            \ENDIF
        \ENDFOR
        \STATE{Choose $y \in Y$ with the largest $U_T(y)$, with ties broken by choosing the $y$ with the largest $D'_T(y)$.}\label{algo:RG:select_y_end}
        \STATE{Modify $T$ by removing the edge between $x$ and its parent and connecting $x$ to $y$.}
        \STATE{Set $\Delta_x = U_T(y) +1-U_T(x)$}\label{algo:RG:update_start}
        \IF{$\Delta_x \ne 0$}
            \STATE{Set $U_T(z) = U_T(z) + \Delta_x, \ \forall z \in T_x(v; T)$}
        \ENDIF
        \STATE{Set $\Delta_T(y) = \max(D_T(y), D_T(x)+1)-D_T(y)$}
        \WHILE{$\Delta_y \ne 0$}
            \STATE{Set $D_T(y) = D_T(y)+\Delta_y$}
            \STATE{Set $\Delta_y = \max(D_T(\parent{y}), D_T(y)+1)-D_T(\parent{y})$}
            \STATE{Set $y=\parent{y}$}
        \ENDWHILE \label{algo:RG:update_end}
    \ENDFOR
\ENDFOR
\RETURN{$T^*_v = T$ and $F^*_v$ computed using \eqref{equ:objective_function_deg}.}
\end{algorithmic}
\end{algorithm}

\section{Simulation Results}\label{sec:simulation_results}
In this section, we present simulation results using both synthetic and real world networks to evaluate the performance of the proposed estimators. We use the following three common centrality measures as benchmarks to compare with our estimators.
\begin{enumerate}[(i)]
  \item The distance centrality of $v \in G$ is defined as
    \begin{align*}
        C_D(v) = \sum_{i \in V_\Exp} d(v, i),
    \end{align*}
    and the node with minimum distance centrality is called the distance center (DC). It is shown in \cite{Shah2011} that the DC is the ML estimator for regular trees under a SI model where all nodes are explicit.
  \item The closeness centrality of $v \in G$ is defined as
    \begin{align*}
        C_C(v) = \sum_{i \in V_\Exp, i \ne v} \frac{1}{d(v, i)},
    \end{align*}
    and the node with maximum closeness centrality is called the closeness center (CC).
  \item The betweenness centrality of $v \in G$ is defined as
    \begin{align*}
        C_B(v) = \sum_{i, j \in V_\Exp, i \ne j \ne v} \frac{\sigma_{ij}(v)}{\sigma_{ij}},
    \end{align*}
    where $\sigma_{ij}$ is the number of shortest paths between node $i$ and node $j$, and $\sigma_{ij}(v)$ is the number of those shortest paths that contain $v$. We call the node with maximum closeness centrality the betweenness center (BC).
\end{enumerate}

\subsection{Tree Networks}
We evaluate the performance of the JCE algorithm on three kinds of synthetic tree networks: regular tree networks where the node degree is randomly chosen from $[3,6]$, and two types of random trees, denoted as random-1 and random-2, where the degree of every node is randomly chosen from $[3,6]$ and $\{3,6\}$ respectively. Note that random-1 trees are less ``regular'' than regular trees, and random-2 trees are even less ``regular''. For each kind of synthetic tree network, we perform 1000 simulation runs. In each simulation run, we randomly generate a tree, and choose a node to be the infection source. Then we simulate the infection using a SI model, where $p$ is chosen uniformly from $(0,1)$ and $q_v$ is chosen uniformly from $[\max(0,2-1/p), 1]$ for each node $v$. The spreading terminates when the number of infected nodes is greater than 200. We then run JCE on the explicit nodes to estimate the infection source and compare the result with the benchmarks.

\begin{figure}[!ht]
  \centering
  \includegraphics[width=0.4\textwidth]{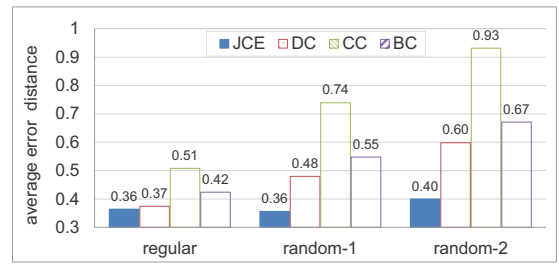}
  \caption{Average error distances for different tree networks. The average diameters of the regular, random-1, and random-2 trees are 14, 13 and 13 hops, respectively.}
\label{fig:trees_result}
\end{figure}

The error distance is the number of hops between the estimated and the actual infection source, and is shown in Fig. \ref{fig:trees_result}. We see that as the underlying network becomes less ``regular'', the error distances of all benchmarks increase, while that of JCE remains relatively stable. In practice, the structures of the underlying tree networks are usually far from regular.

\subsection{General Networks}
We evaluate the performance of the MIQCQP and RG algorithms on three kinds of general networks: synthetic small-world networks \cite{Watts1998}, the western states power grid network of the United States \cite{Watts1998} and a small part of the Facebook network with 4039 nodes \cite{McAuley2012}.

\begin{figure}[!t]
  \centering
  \includegraphics[width=0.4\textwidth]{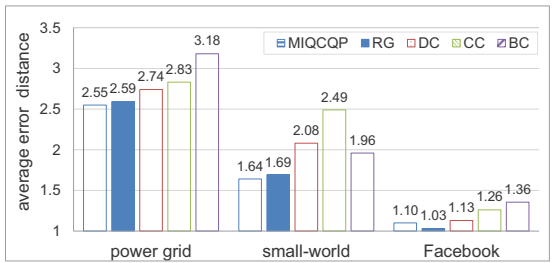}
  \caption{Average error distances for different general networks in the first comparison. The average diameters of the power grid network, small-world network, and Facebook network are 46, 20, and 18 hops, respectively.}
\label{fig:general_networks_setting1_result}
\end{figure}

Since the MIQCQP approach has very high complexity with an average running time about 2000 times that of RG, we restrict the number of infected nodes to be 50 in the first comparison. In each simulation run, the infection probability $p$ is chosen uniformly from $(0,1)$ and $q_v$ is chosen uniformly from $[\max(0,2-1/p), 1]$ for each node $v$. Fig. \ref{fig:general_networks_setting1_result} shows that both MIQCQP and RG have smaller average error distances than all benchmarks, with RG only slightly worse than MIQCQP. Therefore, in the rest of the simulations, we will only compare RG with the benchmarks.

\begin{figure}[!t]
  \centering
	  \subfigure[Power grid network.]{
    \label{fig:error_distance_power_grid}
    \includegraphics[width=0.4\textwidth]{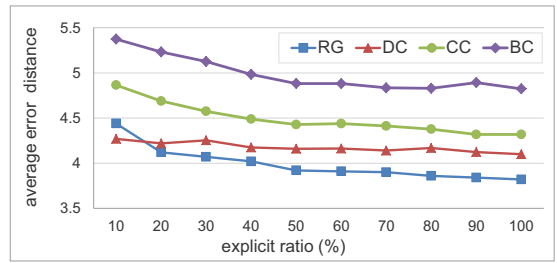}}
  \hspace{0.1cm}
  \subfigure[Small-world network.]{
    \label{fig:error_distance_small_world}
    \includegraphics[width=0.4\textwidth]{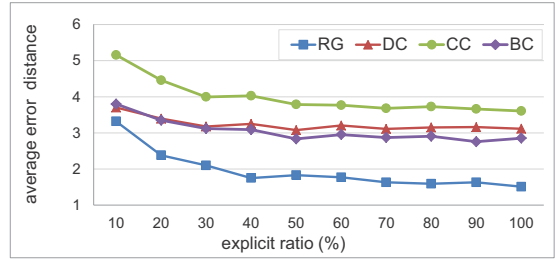}}
  \hspace{0.1cm}
  \subfigure[Facebook network.]{
    \label{fig:error_distance_facebook}
    \includegraphics[width=0.4\textwidth]{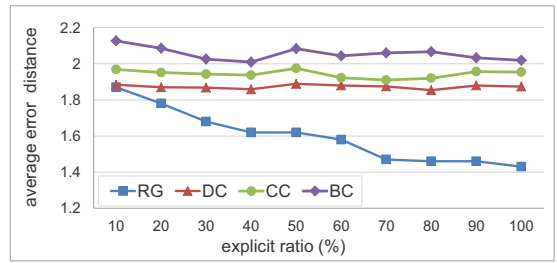}}
  \caption{Average error distances for different explicit ratios in different general networks. The average diameters of the power grid network, small-world network, and Facebook network are 46, 20, and 18 hops, respectively.}
\label{fig:general_networks_setting2_result}
\end{figure}

In the next comparison, we consider infection sizes of more than 200 nodes. In each simulation run, we randomly choose a fraction of the infection nodes to be explicit. We call this fraction the explicit ratio. We perform simulations with the explicit ratio ranging from 10\% to 100\%. Fig. \ref{fig:general_networks_setting2_result} shows the average error distances of RG and all benchmarks for different explicit ratios on all three kinds of general networks. We see that RG has smaller average error distances in almost all cases.

\section{Conclusion}\label{sec:conclusion}
We have derived infection source estimators for a SI model when not all infected nodes can be observed. When the network is a tree, we showed that the estimator is a Jordan center. We proposed an efficient algorithm with complexity $O(n)$ to find the estimator, where $n$ is the size of the network. In the case of general networks, we proposed approximate source estimators based on a MIQCQP formulation, which has high complexity, and a heuristic algorithm with complexity $O(n^3)$. Simulations suggest that our proposed algorithms have better performance than distance, closeness, and betweenness centrality based methods, with the estimated source on average within a few hops of the actual infection source. In this work, we have assumed that we have access only to very limited information. An interesting future direction includes incorporating information like infection times into the source estimator. It is also of interest to investigate source estimation methods for richer models like SIS models when there are only limited observations.

\appendices

\section{Proof of Lemma \ref{lemma:empty_subtree}}\label{appendix:lemma:empty_subtree}

The proof proceeds by mathematical induction on the elapsed time $t$.

\noindent\textbf{Basis step}:
Suppose that $t=1$, and consider any non-observable subtree $T_u(v;G)$. Suppose that $d(v,u) > 1$. Since the infection can spread at most one hop away from $v$ in each time slot, the node $u$ must remain uninfected up to time $t=1$ and the claim holds trivially. We now consider only the neighbors of $v$. Let $u$ be a neighbor of $v$ that is the root of a non-observable subtree, and suppose that every most likely infection path $\tilde{X}^1$ for $(v,t)$ has $\tilde{X}(u,1)=\Inf$. Choose one such most likely infection path $\tilde{X}^1$, and let $X^1$ be another infection path with the same states as $\tilde{X}^1$, except that $X(u,1)=\Sus$. Then, we have
\begin{align*}
\frac{P_v(X^1)}{P_v(\tilde{X}^1)}
&= \frac{1-p}{p(1-q_u)} \geq 1,
\end{align*}
where the last inequality follows from the assumption \eqref{ineq:q}. This is a contradiction, and therefore the claim holds if $t=1$.

\noindent\textbf{Inductive step}:
Suppose that the claim holds for all elapsed times $t \leq n$. Let $t=n+1$, and consider any non-observable subtree $T_u(v;G)$ and a most likely infection path $\tilde{X}^t$ for $(v,t)$ with $\tilde{X}(u,t)=\Inf$.

If $d(v,u) > 1$, let $\tilde{t}$ denote the first infection time of the parent node $r=\parent{u}$ in $\tilde{X}^t$, where $\tilde{t} \geq 1$. Since the infection process follows a SI model, we can treat $r$ as the infection source of $T_{r}(v;G)$, and the remaining elapsed time of the infection process is $t'=t-\tilde{t} \leq n$. From the induction hypothesis, we can construct a most likely infection path $X^{t'}(T_{r}(v;G),[1,t'])$ such that $u$ remains uninfected up to time $t$. The new infection path constructed from $X^t$ by setting $\tilde{X}^{t}(T_{r}(v;G),[\tilde{t}+1,t])=X^{t'}(T_{r}(v;G),[1,t'])$ then has probability at least that of $\tilde{X}^t$, thus proving our claim.

Now suppose that $d(v,u)=1$ and the first infection time of node $u$ in the path $\tilde{X}^t$ is $t_u \in [1,t]$. The previous argument shows that it is possible to choose $\tilde{X}^t$ so that $w$ remains uninfected up to time $t$ for all $w \in T_u(v;G)\backslash\{u\}$ since $d(v,w) > 1$. We now suppose $\tilde{X}^t$ is chosen as such. Let $X^t$ be an infection path that is the same as the most likely infection path $\tilde{X}^t$ but with $X^t(u,\tau) = \Sus$ for all $\tau \leq t$. Then, we have
\begin{align}
P_v(X^t) &= a \prod_{\tau=1}^t  P_v(X^t(u,\tau)=\Sus \mid X^t(u,\tau-1)=\Sus) \nonumber \\
&= a(1-p)^t,  \label{equ:empty_subtree_inductive_P1}
\end{align}
where $a =P_v(X^t(V\backslash T_u(v;G),[1,t]))$. Similarly, we have
\begin{align}
P_v(\tilde{X}^t)
= & a \prod_{\tau = 1}^{t_u-1} \big(P_v(\tilde{X}^t(u,\tau)=\Sus \mid \tilde{X}^t(u,\tau-1)=\Sus) \nonumber\\
& \quad \cdot P_v(\tilde{X}^t(u,\tau)=\Inf \mid \tilde{X}^t(u,\tau-1)=\Sus)\big) \nonumber \\
\cdot \prod_{w \in \children{u}} & \prod_{\tau=t_u+1}^{t} P_v(\tilde{X}^t(w,\tau)=\Sus \mid \tilde{X}^t(w,\tau-1)=\Sus)  \nonumber \\
= & a(1-p)^{t_u-1}p(1-q_u)(1-p)^{(t-t_u)\left|\children{u} \right|} \nonumber \\
\leq &  a(1-p)^{t-1}p(1-q_u),\label{equ:empty_subtree_inductive_P3}
\end{align}
where the last inequality follows because we have assumed that $|\children{u}|\geq 1$. Comparing \eqref{equ:empty_subtree_inductive_P1} and \eqref{equ:empty_subtree_inductive_P3}, and using \eqref{ineq:q}, we obtain $P_v(X^t) \geq P_v(\tilde{X}^t)$, which implies that $X^t$ is also a most likely infection path. Repeating the same argument for all non-observable subtrees with roots that are neighbors of $v$, we obtain the claim, and the lemma is proved.

\section{Proof of Lemma \ref{lemma:better_later}}\label{appendix:lemma:better_later}

Firstly, it is easy to see that every node in $H$ is infected since otherwise, the infection can not reach the leaf nodes of $H$, some of which belong to the set $V_\Exp$. The lower bound in \eqref{equ:first_infection_time_bound} follows because the infection can spread at most one hop away from $v$ in each time slot, and the earliest time for $u$ to be infected is $d(v,u)$. After node $u$ gets infected at time $t_u$, the infection can spread at most $t-t_u$ hops away from $u$. Consider a node $u_l$ such that $d(u,u_l) = \bar{d}(u, T_u(v;H))$. By definition, $u_l \in V_\Exp$. In order for the infection to reach node $u_l$, we require $t-t_u \geq d(u,u_l)$, and \eqref{equ:first_infection_time_bound} is shown.

Next, we show \eqref{equ:optimal_first_infection_time} using mathematical induction on $\bar{d}(v,V_\Exp)$. Note that we do not need to consider the case where $\bar{d}(v,V_{\Exp})=0$, i.e., no non-source nodes exist in $H$.

\noindent\textbf{Basis step}:
Suppose that $\bar{d}(v,V_{\Exp})=1$. Then we have $d(v,u)=1$ for every non-source node $u$ in $H$. We want to show that there exists a most likely infection path so that the first infection time for $u$ is $t_u=t$. If $t=1$, we have $t_u=1$ because both the lower and upper bounds in \eqref{equ:first_infection_time_bound} are equal to 1. We now suppose that $t>1$, and we have $t_u \in [1,t]$ by \eqref{equ:first_infection_time_bound}. Let $\tilde{X}^t$ denote a most likely infection path with $t_u = i$, where $1 \leq i \leq t-1$. We construct another infection path $X^t$ from $\tilde{X}^t$ so that $X^t(V\backslash\{u\},[1,t]) = \tX^t(V\backslash\{u\},[1,t])$ and node $u$ becomes infected only at time $t$. Let $a = P_v(X^t(V\backslash\{u\},[1,t]))$. We then have
\begin{align}
P_v(X^t)
&= a P_v(X^t(T_u(v;G),[1,t])) \nonumber\\
&= (1-p)^{t-1}pq_u.  \label{equ:better_later_basis_P1}
\end{align}
The same derivation as in \eqref{equ:empty_subtree_inductive_P3}, except that here $\tilde{X}^t(u,t)$ is explicit instead of non-observable, gives
\begin{align}
P_v(\tilde{X}^t)
= & a(1-p)^{i-1}pq_u(1-p)^{(t-i)\left| \children{u} \right|} \nonumber \\
\leq &  a(1-p)^{t-1}pq_u,\label{equ:better_later_basis_P2}
\end{align}
where we use the assumption that $| \children{u}|\geq 1$ in the last inequality. By comparing \eqref{equ:better_later_basis_P1} and \eqref{equ:better_later_basis_P2}, we obtain $P_v(X^t) \geq P_v(\tX^t)$, and by repeating the same argument for all neighbors of $v$ in $V_\Exp$, we obtain the claim for the basis step.

\noindent\textbf{Inductive step}: We assume that \eqref{equ:optimal_first_infection_time} holds when $\bar{d}(v,V_{\Exp})\leq n$. Fix any $u \in \children{v}$ and let $m= \bar{d}(u, T_u(v;H))$. Treat node $u$ as the infection source of the subtree $T_u(v;H)$, then by the induction hypothesis, we obtain a most likely infection path $\tX^t$ with \eqref{equ:optimal_first_infection_time} holding for any node $w \in T_u(v;H) \backslash \{u\}$ since $m \leq  \bar{d}(v,V_{\Exp})-1=n$.

For any node $w \in N_1(u; T_u(v;H))$, we have the first infection time of $w$ is $t-(m-1)$ (note that $T_u(v;H)$ is not a non-observable subtree), which implies that $t_u \leq t-m$. We want to show that $\hat{t}_u=t-m$. Suppose that we have $t_u = i$, for some $1 \leq i \leq t-m-1$ in $\tX^t$. Then, using the same arguments as in the basis step, we can construct another infection path with probability at least that of $\tX^t$ but with $t_u = t-m$. This implies our claim, and the lemma is proved.

\bibliography{IEEEabrv,SIS}{}
\bibliographystyle{IEEEtran}

\begin{biography}
[{\includegraphics[width=1in,height =1.25in,clip,keepaspectratio]{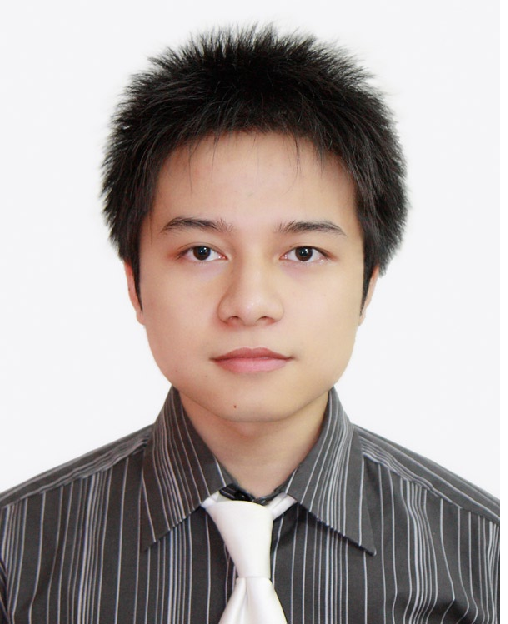}}]
{Wuqiong Luo} (S'12) received the B.Eng.\ degree in electrical and electronic engineering (with first class hons.) from Nanyang Technological University, Singapore, in 2010. He is currently working toward the Ph.D.\ degree in electrical and electronic engineering at Nanyang Technological University.

His research interests are in source estimation and identification in communication networks.

Mr. Luo was coawarded the Best Student Paper Award at the 46th Asilomar Conference on Signals, Systems, and Computers.
\end{biography}

\begin{biography}
[{\includegraphics[width=1in,height =1.25in,clip,keepaspectratio]{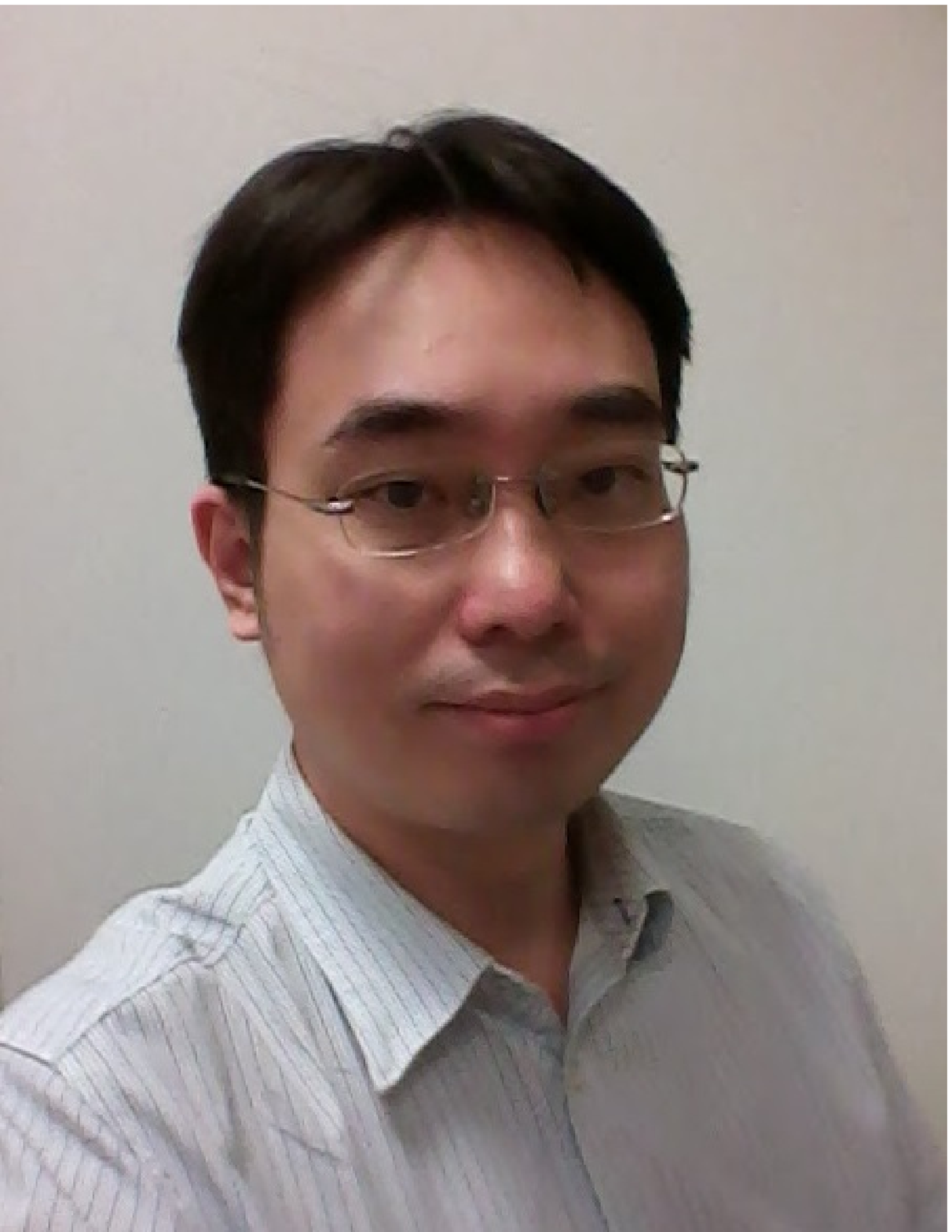}}]
{Wee Peng Tay} (S'06 M'08) received the B.S. degree in Electrical Engineering and Mathematics, and the M.S. degree in Electrical Engineering from Stanford University, Stanford, CA, USA, in 2002. He received the Ph.D. degree in Electrical Engineering and Computer Science from the Massachusetts Institute of Technology, Cambridge, MA, USA, in 2008. He is currently an Assistant Professor in the School of Electrical and Electronic Engineering at Nanyang Technological University, Singapore. His research interests include distributed decision making, data fusion, distributed algorithms, communications in ad hoc networks, machine learning, and applied probability.

Dr. Tay received the Singapore Technologies Scholarship in 1998, the Stanford University President's Award in 1999, and the Frederick Emmons Terman Engineering Scholastic Award in 2002. He is the coauthor of the best student paper award at the 46th Asilomar conference on Signals, Systems, and Computers. He is currently serving as a vice chair of an Interest Group in IEEE MMTC, and has served as a technical program committee member for various international conferences.
\end{biography}

\begin{biography}
[{\includegraphics[width=1in,height =1.25in,clip,keepaspectratio]{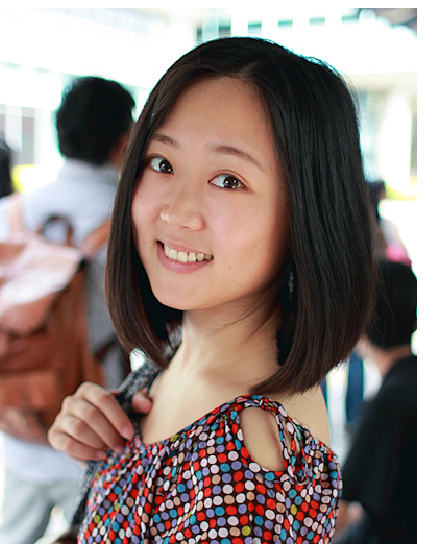}}]
{Mei Leng} (S'07-M'10) received the B.Eng. degree from University of Electronic Science and Technology of China (UESTC), Chengdu, China, in 2005, and the Ph.D. degree from The University of Hong Kong, Hong Kong, in 2011.

She is currently a Research Fellow at the School of Electrical and Electronic Engineering, Nanyang Technological University, Singapore. Her current research interests include statistical signal processing, optimization, machine learning, as well as Bayesian analysis, with applications to wireless sensor networks and wireless communication systems.
\end{biography}

\end{document}